\documentclass[10pt,a4paper,onecolumn]{article}
\usepackage{marginnote}
\usepackage{graphicx}
\usepackage{xcolor}
\usepackage{authblk,etoolbox}
\usepackage{titlesec}
\usepackage{calc}
\usepackage{tikz}
\usepackage{hyperref}
\hypersetup{colorlinks,breaklinks,
            urlcolor=[rgb]{0.0, 0.5, 1.0},
            linkcolor=[rgb]{0.0, 0.5, 1.0}}
\usepackage{caption}
\usepackage{tcolorbox}
\usepackage{amssymb,amsmath}
\usepackage{ifxetex,ifluatex}
\usepackage{seqsplit}
\usepackage{xstring}

\usepackage{fixltx2e} 
\usepackage[
  backend=biber,
]{biblatex}
\bibliography{paper.bib}

\let\textttOrig=\texttt
\def\texttt#1{\expandafter\textttOrig{\seqsplit{#1}}}
\renewcommand{\seqinsert}{\ifmmode
  \allowbreak
  \else\penalty6000\hspace{0pt plus 0.02em}\fi}

\makeatletter
\let\href@Orig=\href
\def\href@Urllike#1#2{\href@Orig{#1}{\begingroup
    \def\Url@String{#2}\Url@FormatString
    \endgroup}}
\def\href@Notdoi#1#2{\def\tempa{#1}\def\tempb{#2}%
  \ifx\tempa\tempb\relax\href@Urllike{#1}{#2}\else
  \href@Orig{#1}{#2}\fi}
\def\href#1#2{%
  \IfBeginWith{#1}{https://doi.org}%
  {\href@Urllike{#1}{#2}}{\href@Notdoi{#1}{#2}}}
\makeatother

\usepackage[top=3.5cm, bottom=3cm, right=1.5cm, left=1.0cm,
            headheight=2.2cm, reversemp, includemp, marginparwidth=4.5cm]{geometry}

\titleformat{\section}
  {\normalfont\sffamily\Large\bfseries}
  {}{0pt}{}
\titleformat{\subsection}
  {\normalfont\sffamily\large\bfseries}
  {}{0pt}{}
\titleformat{\subsubsection}
  {\normalfont\sffamily\bfseries}
  {}{0pt}{}
\titleformat*{\paragraph}
  {\sffamily\normalsize}

\usepackage{fancyhdr}
\makeatletter
\let\ps@plain\ps@fancy
\fancyheadoffset[L]{4.5cm}
\fancyfootoffset[L]{4.5cm}

\definecolor{linky}{rgb}{0.0, 0.5, 1.0}

\newtcolorbox{repobox}
   {colback=red, colframe=red!75!black,
     boxrule=0.5pt, arc=2pt, left=6pt, right=6pt, top=3pt, bottom=3pt}

\newcommand{\ExternalLink}{%
   \tikz[x=1.2ex, y=1.2ex, baseline=-0.05ex]{%
       \begin{scope}[x=1ex, y=1ex]
           \clip (-0.1,-0.1)
               --++ (-0, 1.2)
               --++ (0.6, 0)
               --++ (0, -0.6)
               --++ (0.6, 0)
               --++ (0, -1);
           \path[draw,
               line width = 0.5,
               rounded corners=0.5]
               (0,0) rectangle (1,1);
       \end{scope}
       \path[draw, line width = 0.5] (0.5, 0.5)
           -- (1, 1);
       \path[draw, line width = 0.5] (0.6, 1)
           -- (1, 1) -- (1, 0.6);
       }
   }

\patchcmd{\@maketitle}{center}{flushleft}{}{}
\patchcmd{\@maketitle}{center}{flushleft}{}{}
\patchcmd{\@maketitle}{\LARGE}{\LARGE\sffamily}{}{}
\def\maketitle{{%
  
  \AB@maketitle}}
\makeatletter
\renewcommand\AB@affilsepx{ \protect\Affilfont}
\renewcommand\AB@affilnote[1]{{\bfseries #1}\hspace{3pt}}
\renewcommand{\affil}[2][]%
   {\newaffiltrue\let\AB@blk@and\AB@pand
      \if\relax#1\relax\def\AB@note{\AB@thenote}\else\def\AB@note{#1}%
        \setcounter{Maxaffil}{0}\fi
        \begingroup
        \let\href=\href@Orig
        \let\texttt=\textttOrig
        \let\protect\@unexpandable@protect
        \def\thanks{\protect\thanks}\def\footnote{\protect\footnote}%
        \@temptokena=\expandafter{\AB@authors}%
        {\def\\{\protect\\\protect\Affilfont}\xdef\AB@temp{#2}}%
         \xdef\AB@authors{\the\@temptokena\AB@las\AB@au@str
         \protect\\[\affilsep]\protect\Affilfont\AB@temp}%
         \gdef\AB@las{}\gdef\AB@au@str{}%
        {\def\\{, \ignorespaces}\xdef\AB@temp{#2}}%
        \@temptokena=\expandafter{\AB@affillist}%
        \xdef\AB@affillist{\the\@temptokena \AB@affilsep
          \AB@affilnote{\AB@note}\protect\Affilfont\AB@temp}%
      \endgroup
       \let\AB@affilsep\AB@affilsepx
}
\makeatother

\renewcommand\Affilfont{\sffamily\small\mdseries}
\ifnum 0\ifxetex 1\fi\ifluatex 1\fi=0 
  \usepackage[T1]{fontenc}
  \usepackage[utf8]{inputenc}
  \usepackage{lmodern}

\else 
  \ifxetex
    \usepackage{mathspec}
  \else
    \usepackage{fontspec}
  \fi
  \defaultfontfeatures{Ligatures=TeX,Scale=MatchLowercase}

\fi
\IfFileExists{upquote.sty}{\usepackage{upquote}}{}
\IfFileExists{microtype.sty}{%
\usepackage{microtype}
\UseMicrotypeSet[protrusion]{basicmath} 
}{}

\usepackage{hyperref}
\hypersetup{unicode=true,
            pdftitle={Fruitbat: A Python Package for Estimating Redshifts of Fast Radio Bursts},
            pdfborder={0 0 0},
            breaklinks=true}
\let\addcontentslineOrig=\addcontentsline
\def\addcontentsline#1#2#3{\bgroup
  \let\texttt=\textttOrig\addcontentslineOrig{#1}{#2}{#3}\egroup}
\let\markbothOrig\markboth
\def\markboth#1#2{\bgroup
  \let\texttt=\textttOrig\markbothOrig{#1}{#2}\egroup}
\let\markrightOrig\markright
\def\markright#1{\bgroup
  \let\texttt=\textttOrig\markrightOrig{#1}\egroup}

\usepackage{graphicx,grffile}
\makeatletter
\def\maxwidth{\ifdim\Gin@nat@width>\linewidth\linewidth\else\Gin@nat@width\fi}
\def\maxheight{\ifdim\Gin@nat@height>\textheight\textheight\else\Gin@nat@height\fi}
\makeatother
\setkeys{Gin}{width=\maxwidth,height=\maxheight,keepaspectratio}
\IfFileExists{parskip.sty}{%
\usepackage{parskip}
}{
\setlength{\parindent}{0pt}
\setlength{\parskip}{6pt plus 2pt minus 1pt}
}
\ifx\paragraph\undefined\else
\let\oldparagraph\paragraph
\renewcommand{\paragraph}[1]{\oldparagraph{#1}\mbox{}}
\fi
\ifx\subparagraph\undefined\else
\let\oldsubparagraph\subparagraph
\renewcommand{\subparagraph}[1]{\oldsubparagraph{#1}\mbox{}}
\fi

\title{Fruitbat: A Python Package for Estimating Redshifts of Fast Radio Bursts}

        \author[1,2]{Adam J. Batten}

      \affil[1]{Centre for Astrophysics and Supercomputing, Swinburne University of Technology, PO Box 218, Hawthorn, VIC 3122, Australia}
      \affil[2]{ARC Centre of Excellence for All Sky Astrophysics in 3 Dimensions (ASTRO 3D)}

  \date{\vspace{-5ex}}

\begin{document}
\maketitle

\marginpar{
  \sffamily\small

  {\bfseries DOI:} \href{https://doi.org/10.21105/joss.01399}{\color{linky}{10.21105/joss.01399}}

  \vspace{2mm}

  {\bfseries Software}
  \begin{itemize}
    \setlength\itemsep{0em}
    \item \href{https://github.com/openjournals/joss-reviews/issues/1399}{\color{linky}{Review}} \ExternalLink
    \item \href{https://github.com/abatten/fruitbat}{\color{linky}{Repository}} \ExternalLink
    \item \href{http://dx.doi.org/10.5281/zenodo.2667596}{\color{linky}{Archive}} \ExternalLink
  \end{itemize}

  \vspace{2mm}

  {\bfseries Submitted:} 11 April 2019\\
  {\bfseries Published:} 9 May 2019

  \vspace{2mm}
  {\bfseries License}\\
  Authors of papers retain copyright and release the work under a Creative Commons Attribution 4.0 International License (\href{http://creativecommons.org/licenses/by/4.0/}{\color{linky}{CC-BY}}).
}

\section{Summary}\label{summary}
\texttt{fruitbat} is a Python 2/3 package for estimating redshifts, energies and the galactic dispersion measure contributions of fast radio bursts (FRBs).

FRBs are a class of short duration ($\sim$ 30 ms) transient radio sources of unknown extragalactic origin (Lorimer, Bailes, McLaughlin, Narkevic, \& Crawford (2007), Thornton et al. (2013), Petroff et al. (2015), CHIME/FRB Collaboration et al. (2019)). There is currently at least 65 confirmed FRB detections (Petroff et al. (2016)), while the new CHIME telescope is expected to add of order 10 FRBs per day (Chawla et al. (2017)).

The defining feature of FRBs, and one that sets them apart from other radio transient events, is their extremely high dispersion measure (DM), generated by the integration of free electron column density along the line-of-sight. FRBs have DM values significantly larger than the estimated Milky Way contribution along the line-of-sight can provide, leading many to believe that their origin is extragalactic. The extragalactic origin of FRBs was confirmed with the host galaxy localisation of a repeating FRB (FRB 121102) to a redshift of $z = 0.197$ (Tendulkar et al. (2017)).

Unfortunately most telescopes do not have high enough resolution to unambiguously localise the host galaxies of FRBs and instead an upper limit of their redshifts can be estimated using a DM-redshift relation; typically calculated via analytic means (e.g. Ioka (2003), Inoue (2004), Zhang (2018)).

We have constructed the tool \texttt{fruitbat} to assist the estimation of redshifts and galactic DM values of FRBs. \texttt{fruitbat} generates and utilises ‘look-up tables’ of existing DM-redshift relations found in the literature (Ioka (2003), Inoue (2004), Zhang (2018)) in conjunction with cosmological parameters determined from both the WMAP (Komatsu et al. (2009), Komatsu et al. (2011), Hinshaw et al. (2013)) and Planck missions (Planck Collaboration (2014), Planck Collaboration (2016), Planck Collaboration (2018)).

\texttt{fruitbat} allows the user to independently choose the DM-redshift relation and the cosmological parameters which was typically not an option when using the relations from the literature. Additionally, \texttt{fruitbat} explicitly integrates the entire DM-redshift relation at each redshift instead of assuming an average value across redshifts; an assumption that introduces a 6\% error (see Equation (6) in Zhang (2018)). In Figure 1 we compare the different DM-redshift relations and cosmologies that have been built into \texttt{fruitbat}.

Furthermore, \texttt{fruitbat} has the functionality for users to define their own DM relations, create custom cosmologies (including non-$\Lambda$CDM cosmologies) and generate their own look-up tables. This feature in particular allows for much greater flexibility than existing techniques in the analysis of FRBs as well as providing the option of adding non-analytic DM-redshift relations such as those derived from cosmological simulations.

\begin{figure}[ht]
    \centering
    \includegraphics{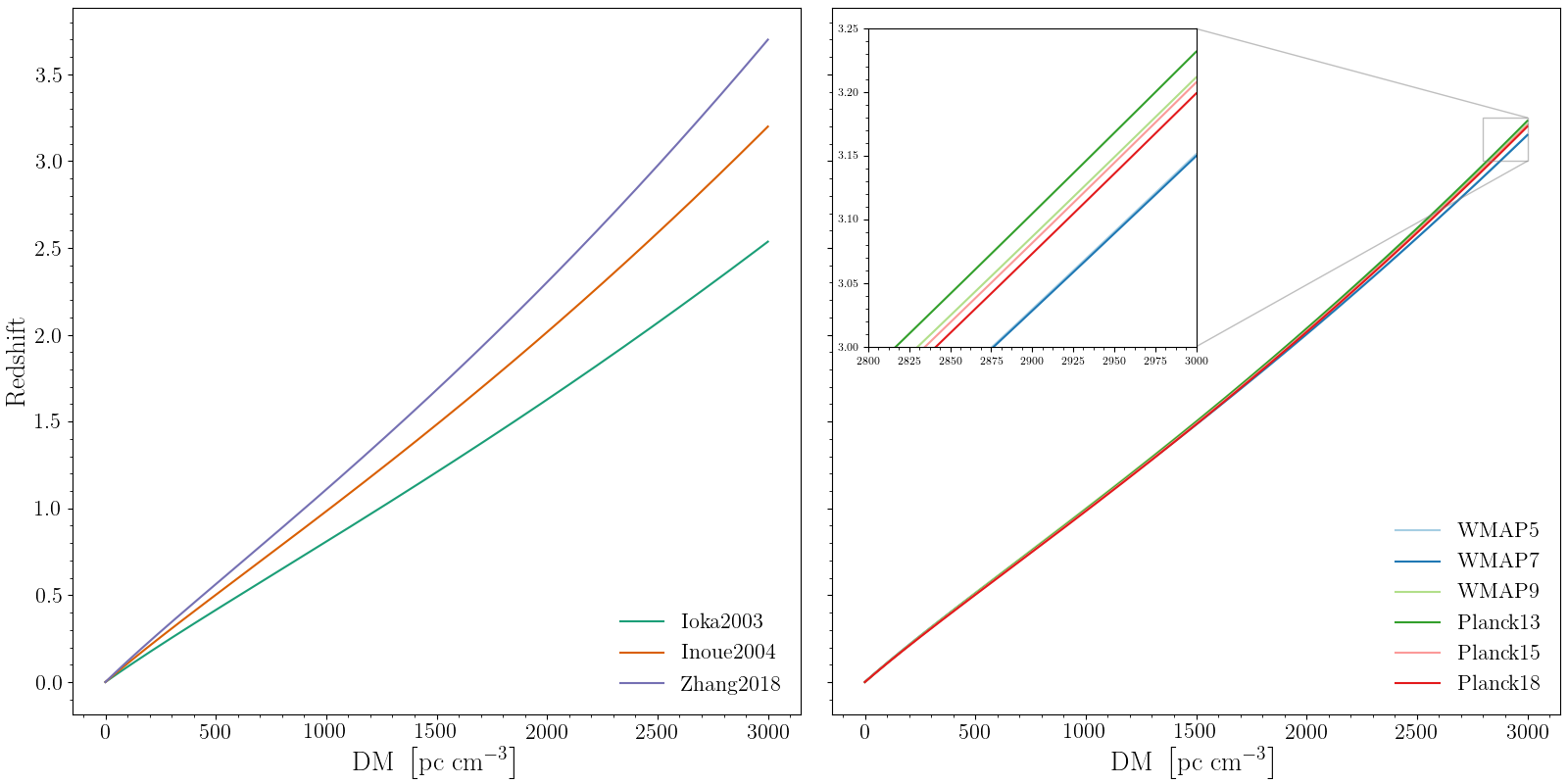}
    \caption{Left: Comparison of three DM-redshift relations assuming a Planck Collaboration (2018) cosmology. Right: Comparison of the Inoue (2004) relation with six different cosmologies.}
    \label{fig:methods_cosmologies}
\end{figure}

\begin{figure}[ht]
    \centering
    \includegraphics{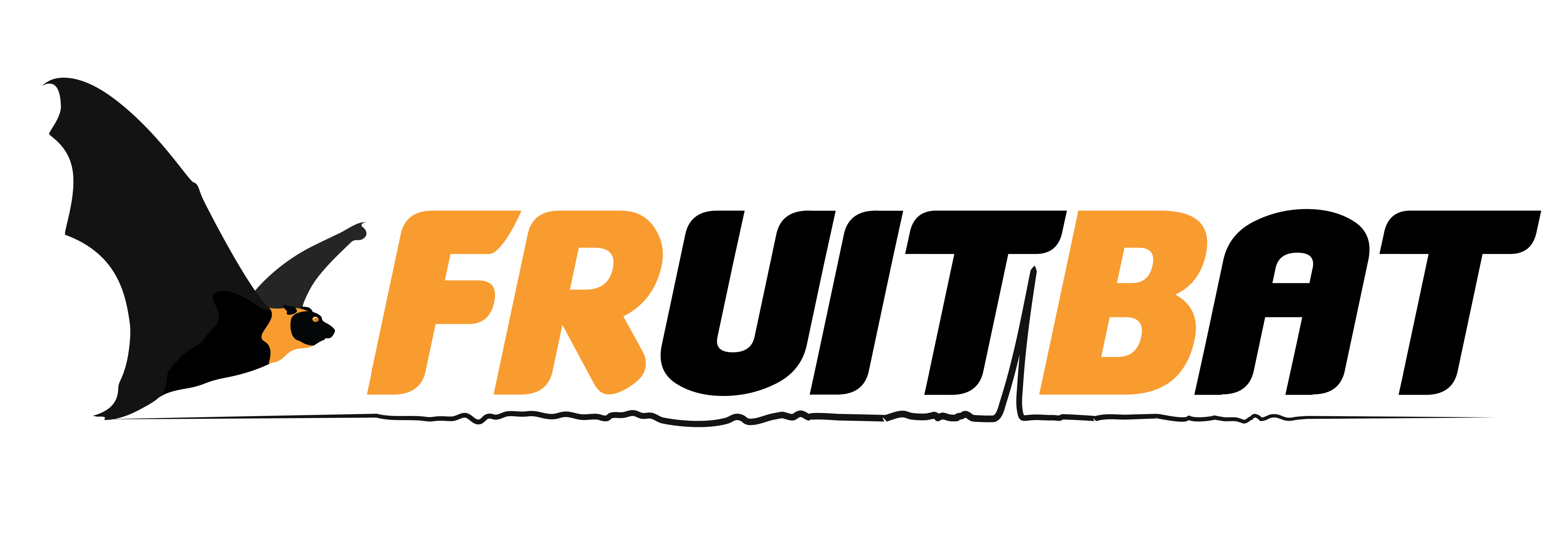}
    \caption{The \texttt{fruitbat} logo.}
    \label{fig:logo}
\end{figure}

To account for the galactic DM contribution due to electrons in the interstellar and circumgalactic medium, \texttt{fruitbat} utilises the YMW16 galactic free electron density model (Yao, Manchester, \& Wang (2017)) to estimate the line-of-sight DM of the Milky Way.

In addition to estimating the redshifts and the line-of-sight galactic DM, \texttt{fruitbat} also has the capability to calculate other quantities of FRBs including: luminosity distances, comoving distance, burst energy and average luminosity.

\texttt{fruitbat} was developed due to the need for a tool that can utilise DM-redshift relations derived analytically and from cosmological simulations (Batten et al. in prep). \texttt{fruitbat} has since been used by Price et al. (MNRAS accepted) to estimate the redshift of FRB 180301.

\texttt{fruitbat} is released under the BSD 3-Clause licence, and is available from PyPi via pip; the \texttt{fruitbat} source code can be found at \href{https://github.com/abatten/fruitbat}{https://github.com/abatten/fruitbat}; tutorials for getting started with \texttt{fruitbat} and online documentation can be found at \href{https://fruitbat.readthedocs.io}{https://fruitbat.readthedocs.io}.

\section{Acknowledgements}\label{acknowledgements}

The author would like to thank Alan Duffy, Ellert van der Velden and Daniel Price for their helpful discussions during the development of this project. The author would also like to thank James Josephides for designing the \texttt{fruitbat} logo. This research was supported by the Australian Research Council Centre of Excellence for All Sky Astrophysics in 3 Dimensions (ASTRO 3D), through project number CE170100013 .

\section*{References}\label{references}

\hypertarget{refs}{}
Chawla, P., Kaspi, V. M., Josephy, A., Rajwade, K. M., Lorimer, D. R., Archibald, A. M., DeCesar, M. E., et al. (2017). A Search for Fast Radio Bursts with the GBNCC Pulsar Survey. \emph{ApJ}, 844, 140. \href{doi:10.3847/1538-4357/aa7d57}{doi:10.3847/1538-4357/aa7d57}

CHIME/FRB Collaboration, Amiri, M., Bandura, K., Bhardwaj, M., Boubel, P., Boyce, M. M., Boyle, P. J., et al. (2019). Observations of fast radio bursts at frequencies down to 400 megahertz. \emph{Nature}, 566, 230–234. \href{doi:10.1038/s41586-018-0867-7}{doi:10.1038/s41586-018-0867-7}

Hinshaw, G., Larson, D., Komatsu, E., Spergel, D. N., Bennett, C. L., Dunkley, J., Nolta, M. R., et al. (2013). Nine-year Wilkinson Microwave Anisotropy Probe (WMAP) Observations: Cosmological Parameter Results. \href{ApJS}, 208, 19. \href{doi:10.1088/0067-0049/208/2/19}{doi:10.1088/0067-0049/208/2/19}

Inoue, S. (2004). Probing the cosmic reionization history and local environment of gammaray bursts through radio dispersion. \emph{MNRAS}, 348, 999–1008. \href{doi:10.1111/j.1365-2966.
2004.07359.x}{doi:10.1111/j.1365-2966.2004.07359.x}

Ioka, K. (2003). The Cosmic Dispersion Measure from Gamma-Ray Burst Afterglows: Probing the Reionization History and the Burst Environment. \emph{ApJ}, 598, L79–L82. \href{doi:10.1086/380598}{doi:10.1086/380598}

Komatsu, E., Dunkley, J., Nolta, M. R., Bennett, C. L., Gold, B., Hinshaw, G., Jarosik, N., et al. (2009). Five-Year Wilkinson Microwave Anisotropy Probe Observations: Cosmological Interpretation. \emph{ApJS}, 180, 330–376. \href{doi:10.1088/0067-0049/180/2/330}{doi:10.1088/0067-0049/180/2/330}

Komatsu, E., Smith, K. M., Dunkley, J., Bennett, C. L., Gold, B., Hinshaw, G., Jarosik, N., et al. (2011). Seven-year Wilkinson Microwave Anisotropy Probe (WMAP) Observations: Cosmological Interpretation. \emph{ApJS}, 192, 18. \href{doi:10.1088/0067-0049/192/2/18}{doi:10.1088/0067-0049/192/2/18}

Lorimer, D. R., Bailes, M., McLaughlin, M. A., Narkevic, D. J., \& Crawford, F. (2007). A Bright Millisecond Radio Burst of Extragalactic Origin. \emph{Science}, 318, 777. \href{doi:10.1126/ science.1147532}{doi:10.1126/ science.1147532}

Petroff, E., Barr, E. D., Jameson, A., Keane, E. F., Bailes, M., Kramer, M., Morello, V., et al. (2016). FRBCAT: The Fast Radio Burst Catalogue. \emph{PASA}, 33, e045. \href{doi:10.1017/pasa.2016.35}{doi:10.1017/pasa.2016.35}

Petroff, E., Johnston, S., Keane, E. F., van Straten, W., Bailes, M., Barr, E. D., Barsdell, B. R., et al. (2015). A survey of FRB fields: limits on repeatability. \emph{MNRAS}, 454, 457–462. \href{doi:10.1093/mnras/stv1953}{doi:10.1093/mnras/stv1953}

Planck Collaboration. (2014). Planck 2013 results. XVI. Cosmological parameters. \emph{A\&A},
571, A16. \href{doi:10.1051/0004-6361/201321591}{doi:10.1051/0004-6361/201321591}

Planck Collaboration. (2016). Planck 2015 results. XIII. Cosmological parameters. \emph{A\&A},
594, A13. \href{doi:10.1051/0004-6361/201525830}{doi:10.1051/0004-6361/201525830}

Planck Collaboration. (2018). Planck 2018 results. VI. Cosmological parameters. arXiv e-prints, arXiv:1807.06209.


Tendulkar, S. P., Bassa, C. G., Cordes, J. M., Bower, G. C., Law, C. J., Chatterjee, S., Adams, E. A. K., et al. (2017). The Host Galaxy and Redshift of the Repeating Fast Radio Burst FRB 121102, \emph{ApJ}, 834, L7. \href{doi:10.3847/2041-8213/834/2/L7}{doi:10.3847/2041-8213/834/2/L7}

Thornton, D., Stappers, B., Bailes, M., Barsdell, B., Bates, S., Bhat, N. D. R., Burgay,
M., et al. (2013). A Population of Fast Radio Bursts at Cosmological Distances. \emph{Science},
341, 53–56. \href{doi:10.1126/science.1236789}{doi:10.1126/science.1236789}

Yao, J. M., Manchester, R. N., \& Wang, N. (2017). A New Electron-density Model for
Estimation of Pulsar and FRB Distances. \emph{ApJ}, 835, 29. \href{doi:10.3847/1538-4357/835/1/29}{doi:10.3847/1538-4357/835/1/29}

Zhang, B. (2018). Fast Radio Burst Energetics and Detectability from High Redshifts.
\emph{ApJ}, 867, L21. \href{doi:10.3847/2041-8213/aae8e3}{doi:10.3847/2041-8213/aae8e3}
\end{document}